\documentclass[aps,prb,twocolumn,superscriptaddress,showpacs]{revtex4-1}

\usepackage{graphicx}
\usepackage{calc}
\usepackage{bm}
\usepackage{color}

\begin{document}

\title{Magnon modes as a joint effect of  surface ferromagnetism and spin-orbite coupling in CoSi chiral topological semimetal}

\author{V.D.~Esin}
\affiliation{Institute of Solid State Physics of the Russian Academy of Sciences, Chernogolovka, Moscow District, 2 Academician Ossipyan str., 142432 Russia}
\author{A.V.~Timonina}
\affiliation{Institute of Solid State Physics of the Russian Academy of Sciences, Chernogolovka, Moscow District, 2 Academician Ossipyan str., 142432 Russia}
\author{N.N.~Kolesnikov}
\affiliation{Institute of Solid State Physics of the Russian Academy of Sciences, Chernogolovka, Moscow District, 2 Academician Ossipyan str., 142432 Russia}
\author{E.V.~Deviatov}
\affiliation{Institute of Solid State Physics of the Russian Academy of Sciences, Chernogolovka, Moscow District, 2 Academician Ossipyan str., 142432 Russia}

\date{\today}

\begin{abstract}
CoSi single crystal is a known realization of a chiral topological semimetal with simultaneously broken mirror and inversion symmetries. In addition to the symmetry-induced spin-orbit coupling, surface ferromagnetism is known in  nominally diamagnetic CoSi structures,  which  appears due to the distorted bonds  and ordered vacancies near the surface. We experimentally investigate electron transport through a thin CoSi flake at high current density.
Surprisingly, we demonstrate $dV/dI(I)$ curves which are qualitatively similar to ones for ferromagnetic multilayers with characteristic $dV/dI$ magnon peaks and  unconventional magnetic field  evolution of the peaks' positions. We understand these observations as a result of current-induced spin polarization due to the significant spin-orbit coupling in CoSi. Scattering of non-equilibrium spin-polarized carriers within the surface ferromagnetic layer is responsible for the precessing spin-wave excitations, so the observed  magnon modes are the joint effect of  surface ferromagnetism and spin-orbit coupling in a CoSi chiral topological semimetal. Thus, thin CoSi flakes behave as   magnetic conductors  with broken inversion symmetry, which is important   for different spintronic phenomena.
\end{abstract}

\maketitle

\section{Introduction}

Topological semimetals is a new and growing field of the condensed matter physics~\cite{armitage}.  Dirac semimetals are characterized by the special points of Brillouin zone with three dimensional linear dispersion. In Weyl semimetals, by breaking inversion  or  time reversal symmetries,  every Dirac point splits  into two Weyl nodes with opposite chiralities.  First experimentally investigated Weyl semimetals  were non-centrosymmetric crystals~\cite{das16,feng2016} like TaAs, WTe$_2$, and MoTe$_2$ with significant spin-orbit coupling.   Also, there are several magnetically-ordered Weyl candidates~\cite{mag1,mag2,mag3,mag4}.

Chiral topological semimetals~\cite{bernevig,zhang} are the natural generalization of Weyl semimetals, they are characterized by simultaneously broken mirror and inversion symmetries.   Bulk band structure and  extremely long surface Fermi arcs have been experimentally  confirmed~\cite{long,cosi1,cosi2,maxChern} for the CoSi  crystal family.

Despite the diamagnetic bulk,  small CoSi single crystals  demonstrate surface ferromagnetism in addition to the symmetry-induced spin-orbit coupling~\cite{li,han}.   The transition metal (Co) d-orbital electron spin up and spin down populations become asymmetric from the exchange interactions near the CoSi surface,  providing the mechanism for the observed surface magnetization~\cite{seo,tai}.

Recently, significant interest appears to magnetic conductors with broken inversion symmetry, which are considered  for different spintronic phenomena~\cite{Edelstein1990,Jungwirth,Silov,Kato,Wunderlich}. The physics of the the spin-orbit torques is similar to one in ferromagnetic multilayers~\cite{myers,tsoi1,tsoi2,katine,single,balkashin,balashov}. In multilayers, the spin-polarized current  transfers part of its  angular momentum from the fixed  to the free ferromagnetic layers (spin torque), which determine the multilayer resistance. In a system lacking  inversion symmetry, the inverse spin galvanic effect (Edelstein effect~\cite{Edelstein1990}) is the electrical generation of  spin density when a current flows~\cite{Jungwirth,Silov,Kato,Wunderlich}. Thus, the spin-orbit coupling serves as a source of polarized spins~\cite{zhelezny} instead of the fixed reference layer in multilayers.  Due to the spin-dependent scattering within the ferromagnet, spin-orbit induced non-equilibrium spin polarization leads, e.g., to the spin-orbit torques and unidirectional magnetoresistance~\cite{zhelezny}.  

Experimentally, spin-orbit torques have been demonstrated in magnetically doped inversion-asymmetric  semiconductor (Ga,Mn)As~\cite{Chernyshov,Fang,Kurebayashi}. On the other hand, coexistence of the surface ferromagnetism and  spin-orbit coupling allows a search for the current-induced magnetization effects also in a CoSi chiral topological semimetal. 
 
Here, we experimentally investigate  electron transport through a thin CoSi flake at high current density.  Surprisingly, we demonstrate $dV/dI(I)$ curves which are qualitatively similar to ones for ferromagnetic multilayers with characteristic $dV/dI$ magnon peaks. The peaks' positions show unconventional magnetic field evolution, they go to zero current in high magnetic field. We understand these observations as a joint effect of  surface ferromagnetism and spin-orbit coupling in CoSi chiral topological semimetal, where scattering of the spin-orbit induced non-equilibrium spins within the surface ferromagnetic layer is responsible for the precessing spin-wave excitations.  Thus, thin CoSi flakes behave as   magnetic conductors  with broken inversion symmetry, which is important   for different spintronic phenomena.

\section{Samples and technique}

\begin{figure}
\includegraphics[width=1\columnwidth]{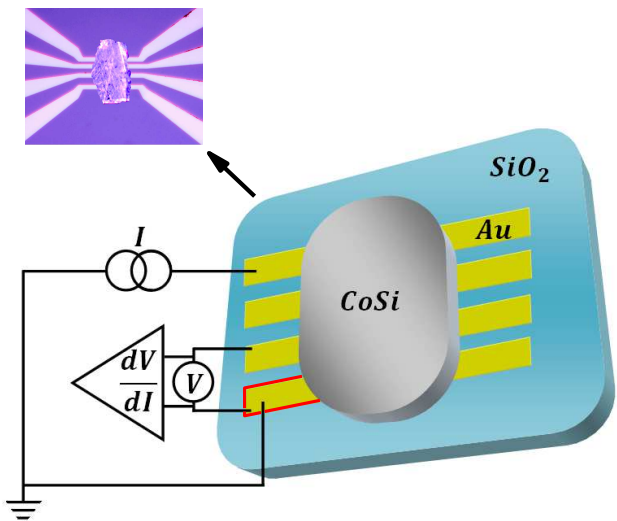}
\caption{(Color online). Sketch of a  sample with electrical connections.  A small (about 100~$\mu\mbox{m}$ size and 0.5~$\mu\mbox{m}$ thick) single-crystal CoSi ﬂake is weakly pressed on the insulating SiO$_2$ substrate with 100 nm thick, 10~$\mu\mbox{m}$ wide Au leads.  Differential resistance of a single Au-CoSi junction is measured in a three-point technique. All the wire resistances are excluded from the circuit. Inset shows an optical image of a typical sample.
}
\label{sample}
\end{figure}

The initial CoSi material was synthesized from cobalt and silicon powders by 10$^{\circ} C/h$ heating in evacuated silica ampoules up to 950$^{\circ} C$. The ampoules were held at this temperature for two weeks and then cooled down to room temperature at 6$^{\circ} C/h$ rate. The obtained material was identified as CoSi with some traces of SiO$_2$ by X-ray analysis. Afterward, CoSi single crystals are grown from this initial load by iodine transport in evacuated silica ampoules at 1000$^{\circ}$. X-ray diffractometry demonstrates cubic structure of the crystals, also, X-ray spectral analysis confirms equiatomic ratio of Co and Si in the composition, without any SiO$_2$ traces.

Thin flakes can be easily obtained from the initial CoSi single crystal by a regular mechanical cleaving  method~\cite{cdas,cosns,black,timnal}. We determine the cleaving surface as (001) one from standard magnetoresistance  measurements~\cite{magres}. CoSi ferromagnetism is a surface effect, so it has  been experimentally demonstrated for thin CoSi structures~\cite{li,han} with high relative thickness of the surface layers. For this reason, we use thin single-crystal CoSi flakes with two $<$0.5~$\mu\mbox{m}$ spaced (001) surfaces to study joint effect of the spin-orbit coupling and ferromagnetism in a chiral CoSi semimetal. 

 A thin CoSi flake is transferred to the metallic contacts, which are  pre-defined on the insulating SiO$_2$ substrate, see  Fig.~\ref{sample}. The contacts pattern consists from 10~$\mu$m wide gold leads, they are formed by lift-off technique after thermal evaporation of 100~nm Au. Van der Waals forces are too weak to hold even a thin CoSi flake on the metallic contacts, thus, it is pressed  slightly with another oxidized silicon substrate. The pressure is removed afterward, so bonding wires can be connected to the Au leads, see inset to Fig.~\ref{sample}.  This procedure provides~\cite{cdas,cosns,timnal} transparent Au-CoSi junctions, stable in different cooling cycles.

The effects of non-equilibrium spin-orbit induced spin polarization are determined by electrical current density, which is a maximum in the vicinity of Au-CoSi contact. For this reason, we investigate differential resistance of a single Au-CoSi junction by a three-point technique, see Fig.~\ref{sample}: the studied contact is grounded, two other contacts are employed to apply current $I$ and measure voltage $V$, respectively. All the wire resistances are excluded  in Fig.~\ref{sample}. In a three-point technique, the measured potential $V$ reflects in-series connected resistances of the grounded contact and some part of the bulk CoSi flake, see Fig.~\ref{sample}.

 To obtain $dV/dI(I)$ characteristics,  the dc current (within $\pm$3~mA, current density corresponds to $\pm10^5$~A/cm$^2$ range)  is additionally modulated by a low ac component  (~5~$\mu\mbox{A}$, $f=7.7$~kHz). We measure both dc (V) and ac (which is proportional to $dV/dI(I)$) voltage components with a dc voltmeter and a lock-in, respectively. The measurements are performed in a  standard 4~K helium cryostat for two different magnetic field orientations. Similar results are obtained from different samples of every type in several cooling cycles.

\begin{figure}
\includegraphics[width=\columnwidth]{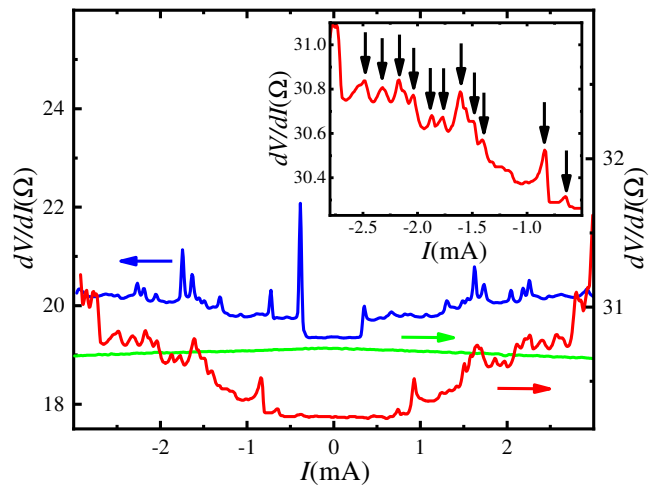}
\caption{(Color online) Typical examples of low-temperature (4.2~K) $dV/dI(I)$ characteristics for two different thin ($\sim$0.5~$\mu$m) CoSi flakes  in zero magnetic field, as depicted by the red and blue curves. The curves shows qualitatively similar behavior:  there are pronounced $dV/dI(I)$ peaks for both current polarities in addition to the overall symmetric increase in $dV/dI(I)$. The peaks are  symmetric with respect to the current sign, they are of different size, see also an enlarged view in the inset (arrows indicate the $dV/dI$ peaks' positions). In addition, green curve shows strictly flat Ohmic $dV/dI(I)$ curve for the reference  2~$\mu$m thick CoSi flake.     Similar  $dV/dI$ peaks are usually connected with spin-wave excitation modes (magnons) in ferromagnetic multilayers~\cite{myers,tsoi1,tsoi2,katine,single,balkashin,balashov}. 
 }
\label{VAC}
\end{figure}

\section{Experimental results}

\begin{figure}
\includegraphics[width=\columnwidth]{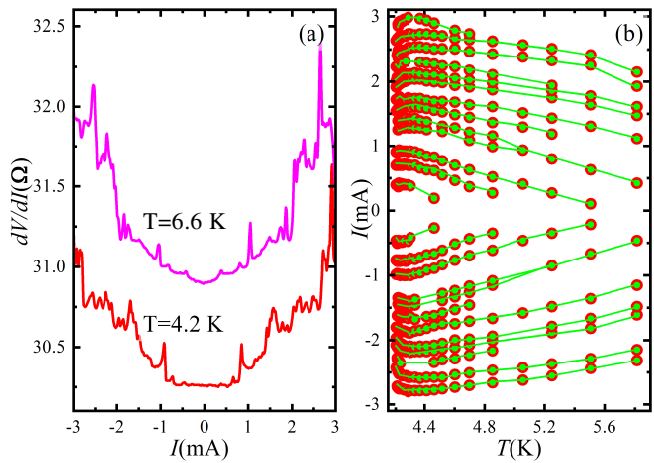}
\caption{(Color online) (a) Temperature dependence of $dV/dI(I)$ curves. The junction resistance is increased with temperature, the most pronounced peaks are shifted to lower currents. (b)  The detailed temperature evolution of the peaks' positions. Any peak demonstrates gradual displacement to lower currents, the low-current peaks are suppressed al lower temperatures than the high-current ones. There is no any noticeable temperature dependence below 4.2~K, while the peaks are fully suppressed above 8~K.  The data are obtained in zero magnetic field.}
\label{TempDepend}
\end{figure}

 Fig.~\ref{VAC} demonstrates typical examples of low-temperature $dV/dI(I)$ characteristics for two Au-CoSi junctions of  different resistance (20 and 31 Ohms, respectively) in zero magnetic field. The curves are obtained for two  different samples, however, they show qualitatively similar behavior:  there are  $dV/dI$ peaks for both current polarities in addition to the overall symmetric increase in $dV/dI(I)$. The peaks are of different size in Fig.~\ref{VAC}, see also an enlarged view in the inset to this figure, they are  symmetric with respect to the current sign. All $dV/dI$ features are well reproducible in different cooling cycles, see also Fig.~\ref{Magnet} below, where data for two field orientations are obtained in separate coolings.  There is no noticeable hysteresis with the current sweep direction for experimental  $dV/dI(I)$ curves. In addition, Fig.~\ref{VAC} shows strictly flat Ohmic $dV/dI(I)$ curve for 2~$\mu$m thick CoSi flake, so  $dV/dI$ features are indeed related to the surface effects.

$dV/dI(I)$ curves are similar to well-known ones for the ferromagnetic multilayers~\cite{myers,tsoi1,tsoi2,katine,single,balkashin,balashov}. The overall symmetric increase in $dV/dI(I)$ is    inconsistent with trivial impurity or roughness scattering at the interface. Instead, it is usually associated with spin-dependent scattering in the magnetically ordered materials. In this case,  sharp $dV/dI$ peaks are usually connected with spin-wave excitation modes (magnons)~\cite{myers,tsoi1,tsoi2,katine,single,balkashin,balashov}: if the spin-transfer effect does not lead to a full magnetization reversal, precessing spin-wave excitations will take place instead, which appears as  sharp peaks in differential resistance.

Below, we analyze temperature or magnetic field evolution of the the $dV/dI$ peaks' positions.   
 
Fig.~\ref{TempDepend} (a) demonstrates temperature dependence of $dV/dI(I)$ curves. We do not observe any noticeable temperature dependence below 4.2~K, while the peaks are fully suppressed above 8~K. For intermediate temperatures, the junction resistance is increasing with temperature, $dV/dI$ peaks go to  lower currents.   The detailed evolution of the peaks' positions can be seen in Fig.~\ref{TempDepend} (b). Both the pronounced and small peaks are perfectly stable, so  we trace all the peaks' positions  until they can be reliably determined in Fig.~\ref{TempDepend} (b).

Any peak demonstrates gradual displacement to lower currents, the low-current peaks are suppressed al lower temperatures than the high-current ones. The continuous temperature evolution of a particular $dV/dI$ peak indicates that any $dV/dI$ peak represents a specific magnon branch~\cite{cosns,timnal}.  

Magnetic field evolution of these magnon branches is shown in Fig.~\ref{Magnet} for two different orientations of the field. First of all, the high-field  behavior is nearly equal in normal and parallel to the Au-CoSi interface magnetic field, so the corresponding magnon branches are definitely not connected with  two-dimensional CoSi surface states. We still observe some effect of the field orientation for low-current (below 1~mA) branches, which could point to their two-dimensional origin~\cite{timnal}.

Secondly, Fig.~\ref{Magnet} shows the magnetic field behavior which  is unconventional for  magnons  in ferromagnetics. The low-current peaks are independent of the magnetic field. The peaks' positions show complicated behavior for intermediate currents and in low magnetic fields, even coalescence of some branches can be seen here. High-current branches demonstrate monotonous decreasing with magnetic field.  This behavior differs significantly from the known direct proportionality of the peaks' positions to the value of magnetic field in ferromagnetic multilayers~\cite{myers,tsoi1,tsoi2,katine,single,balkashin,balashov}.

\begin{center}
\begin{figure}
\includegraphics[width=\columnwidth]{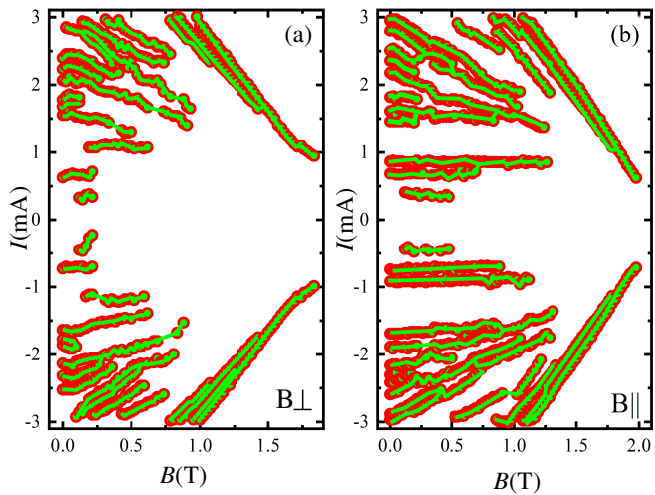}
\caption{
(Color online) Evolution of $dV/dI(I)$ peaks' positions  for two different orientations of magnetic field.  Magnetic field is normal to the CoSi flake plane (a) or it  is oriented parallel to the Au-CoSi interface (b). The high-field behavior is nearly equal in these two configurations, but is unconventional for ferromagnetic multilayers~\cite{myers,tsoi1,tsoi2,katine,single,balkashin,balashov}.  We still observe some effect of the field orientation for low-current (below 1~mA) branches, which could point to their two-dimensional origin~\cite{timnal}. For intermediate currents, even coalescence of some branches can be seen here.  The data are presented for 4.2~K temperature.
}
\label{Magnet}
\end{figure}
\end{center}

\section{Discussion} \label{disc}

As a result, we observe the $dV/dI(I)$ curves with multiple pronounced peaks, which are similar to spin-wave excitations in magnetic ordered structures, like ferromagnetic multilayers~\cite{myers,tsoi1,tsoi2,katine,single,balkashin,balashov}, but the peaks' magnetic field evolution is unusual. 

For multilayers, magnon exitations appear as a result of spin-dependent scattering of spin-polarized current between fixed and free ferromagnetic layers~\cite{myers}. For magnetic Weyl semimetals~\cite{timnal},  spin-polarized topological surface state represents the free layer, while ferromagnetic bulk  serves as a source of spin-polarized carriers (fixed layer). 

We can expect similar physics for CoSi, despite nominally there is no magnetic layers in Au-CoSi structure. 

(i)  Surface ferromagnetism~\cite{li,han} takes a role of a reference layer, where spin-wave exitations can be induced by spin-polarized current. Ordered vacancies~\cite{seo} and  distorted bonds near the surface~\cite{tai} provide the mechanism for the  surface ferromagnetism.

(ii)  Spin-orbit coupling serves as a source of polarized spins instead of the fixed reference layer~\cite{zhelezny}.  It can be the  bulk inverse spin-galvanic effect~\cite{Edelstein1990,Jungwirth,Silov,Kato,Wunderlich} or spin polarization within the topological  surface states~\cite{timnal}.  In the latter case, one could expect  pronounced dependence on the direction (in-plane or normal) of the magnetic field~\cite{timnal}. In our experiment, the high-field evolution is nearly equal in normal and parallel to the Au-CoSi interface magnetic fields. Thus, spin polarization due to the bulk Edelstein effect seems to be dominant in high magnetic fields, where topological  surface states are destroyed. The low-field behavior is more sophisticated for low-current branches, so we can not exclude the combined origin of spin-polarization in low fields.

 Despite the observed $dV/dI$ peaks are similar to those obtained in  ferromagnetic multilayers, their evolution with temperature and magnetic field is unusual in Figs.~\ref{TempDepend} and ~\ref{Magnet}.  In general, the $dV/dI$ peak  position $I_{sw}$ is described by Slonczewski model~\cite{slonczewski,katine}. Slightly simplified,
 \begin{equation}
I_{sw}(B) \sim \alpha \gamma e \sigma B, \label{eq}
\end{equation}
where $\alpha$ is the damping parameter, $\gamma$ is the gyromagnetic ratio, $\sigma$ is the total spin of the free layer. 

In multilayers the total spin $\sigma$ is a constant due to the monodomain regime of low-size structures, so $I_{sw}$ is linearly increasing~\cite{myers,katine,cosns} with the magnetic field  $B$.  In contrast, the size of ferromagnetic domains is typically much smaller than  the 5~$\mu$m distance between the gold leads in our samples. Thus, the domain wall regions should be important for spin-dependent scattering within the conducting CoSi flake, so the domain walls determine the total spin  $\sigma$ in our case. Increasing of the magnetic field  removes domain walls between the Au leads, so $\sigma$ goes to zero in higher fields. This is the origin of the unconventional evolution of magnon modes with magnetic field in Fig.~\ref{Magnet}. Because of variously oriented spins in the domain wall regions, the effect is not sensitive to the mutual orientation of the magnetic field, current, and sample crystallographic axes, as we observe in Fig.~\ref{Magnet} (a) and (b).

The relative strength of spin-orbit coupling in  CoSi is of the order of millielectronvolts~\cite{Burkovetal2018}. This explains the monotonous $I_{sw}(T)$ dependence in Fig.~\ref{TempDepend} (b): spin-orbit-induced spin polarization is diminishing with temperature, so $dV/dI$ peaks disappear due to the temperature smearing above 8~K.

\section{Conclusion}

As a conclusion, CoSi single crystal is a known realization of a chiral topological semimetal with simultaneously broken mirror and inversion symmetries. In addition to the symmetry-induced spin-orbit coupling, surface ferromagnetism is known in  nominally diamagnetic CoSi structures,  which  appears due to the distorted bonds  and ordered vacancies near the surface. We experimentally investigate electron transport through a thin CoSi flake at high current density.
Surprisingly, we demonstrate $dV/dI(I)$ curves which are qualitatively similar to ones for ferromagnetic multilayers with characteristic $dV/dI$ magnon peaks and  unconventional magnetic field  evolution of the peaks' positions. We understand these observations as a result of current-induced spin polarization due to the significant spin-orbit coupling in CoSi. Scattering of non-equilibrium spin-polarized carriers within the surface ferromagnetic layer is responsible for the precessing spin-wave excitations, so the observed  magnon modes is  a joint effect of  surface ferromagnetism and spin-orbit coupling in a CoSi chiral topological semimetal. Thus, thin CoSi flakes behave as   magnetic conductors  with broken inversion symmetry, which is important   for different spintronic phenomena.

\acknowledgments

We wish to thank Yu.S.~Barash and V.T.~Dolgopolov for fruitful discussions, S.S~Khasanov for X-ray sample characterization.  We gratefully acknowledge financial support partially by the RFBR  (project No.~19-02-00203),  and RF State task.

\end{document}